\documentclass[12pt,aps,prb,preprint,draft]{revtex4}

\usepackage{amsmath}
\usepackage{graphicx}

\newcommand{\kms}{\ensuremath{\hbox{km}\cdot \hbox{s}^{-1}}}

\newcommand{\bfr}{{\bf r}}

\newcommand{\spose}[1]{\hbox to 0pt{#1\hss}}
\newcommand{\lta}{\mathrel{\spose{\lower
3pt\hbox{$\mathchar"218$}}
\raise 2.0pt\hbox{$\mathchar"13C$}}}

\begin{document}

\title{Superluminal apparent motions in distant radio sources}

\author{Micha{\l} J.\ Chodorowski}
\email{michal@camk.edu.pl} 
\affiliation{Copernicus Astronomical Center, Bartycka 18, 00--716 
Warsaw, Poland}

\begin{abstract}
We derive the prediction of the standard model of superluminal
radio sources for the apparent transverse velocity of a radio
source located at redshift $z$. The apparent velocity of the
source is reduced by a factor of $1 + z$ compared to that of a
similar nearby source. The cause of this reduction is recession of
the distant source due to the expansion of the universe. The
apparent velocity of a source can be estimated from its redshift
and proper motion using the values of the Hubble constant and the
mean densities of different energy components in the universe. We
derive an expression for the velocity valid for the currently
favored cosmological model: a flat universe with a nonzero
cosmological constant.
\end{abstract}

\maketitle

\section{Introduction}
\label{sec:intro}
In 1966, Martin Rees predicted that ``an object moving
relativistically in suitable directions may appear to a distant
observer to have a transverse velocity much greater than the velocity
of light.''\cite{rees} A few years later, such motion was discovered
in very distant astronomical radio sources such as radio galaxies and
quasars. The motion and the sources where it took place are called
superluminal, that is, faster than light. The discovery was the
spectacular result of a new technique known as very long baseline
interferometry (VLBI). This technique has enabled the mapping of the
morphologies of radio sources to accuracy better than milliseconds of
arc.

Many radio galaxies and quasars contain in
their nuclei compact sources of radio emission with several
components that appear to move apart in successive VLBI images.
Their apparent transverse velocity of separation often exceeds the
speed of light. Superluminal motion has been observed in over a
hundred of sources.\cite{keller} However, it is no unique to radio
galaxies and quasars. There also is a class of recently discovered
galactic superluminal sources called microquasars. All of
these sources (galactic and extragalactic) are thought to
contain a black hole, which is responsible for the ejection of mass
at high velocities.

Superluminal motion does not contradict special relativity. In the
generally accepted standard model, it can be explained as a light
travel-time effect. Superluminal radio sources can be modeled by
one or more radiating ``blobs,'' moving at a relativistic velocity
away from a stationary ``core.'' Imagine a blob of matter starting
at the core and moving toward an observer very fast and nearly
head-on. When the blob is at the core, it emits some light toward
the observer. After it has moved toward the observer (and slightly
to the side), it again emits light toward the observer. Because it
is closer to the observer, this light will take a shorter time to
travel to the observer. If we ignore this fact, then we will 
underestimate the true time interval, and so we will overestimate
the speed.

More quantitatively, the observed transverse velocity of the
separation of the blob from the core, $v_a$, is related to the
true velocity, $v$, and the angle to the line of sight, $\theta$, by
\begin{equation}
v_a = \frac{v \sin\theta}{1 - (v/c)\cos\theta} , 
\label{eq:v_a_stat}
\end{equation}
where $c$ is the velocity of light. The angle $\theta = 0$
corresponds to motion directly toward the observer; $\theta = \pi$
corresponds to motion directly away from the observer.
Equation~(\ref{eq:v_a_stat}) shows that if the motion is
relativistic, $v \lta c$, and almost in the direction of the
observer, separation at superluminal apparent velocities
will be observed.

The derivation of Eq.~(\ref{eq:v_a_stat}) can be found in many
textbooks on astrophysics (see, for example,
Refs.~\onlinecite{krolik,shu}). To our knowledge, however, none of
them stresses its fundamental limitation: it is valid only for nearby
sources. However as we will explain, most sources are actually very
distant. The apparent transverse velocity of a distant radio source
also depends on its redshift, defined as the relative difference
between the observed and the rest wavelength of the emitted light, $z
\equiv (\lambda_{\rm obs} - \lambda_{\rm em})/\lambda_{\rm
em}$.\cite{redshift} Specifically, it is reduced by a factor of $1 +
z$.\cite{pz87} However, this effect is only briefly
discussed,\cite{cite} and its physical interpretation appears to be
missing in the literature. In this paper, we derive the effect from
first principles and elucidate its physical meaning.

The proper motion of a celestial object is the change in time of its
angular position on the sky. To estimate the apparent velocity of an
object with proper motion $\mu$, located at the redshift $z$, we need
its angular diameter distance, $D_A$. $D_A$ depends on the redshift as
well as on various cosmological parameters. To date, all analyses of
superluminal radio sources at large distances employ the classical
Mattig formula for $D_A$, which is valid only for a vanishing
cosmological constant, $\Lambda$.\cite{lambda} However, cosmological
observations suggest that $\Lambda \simeq 0.7$ in units of the
critical density.\cite{value} We will derive an expression in integral
form for $D_A$ for $\Lambda \neq 0$, corresponding to a flat
universe. Following Pen,\cite{pen} we also obtain its analytic
approximation.

The paper is organized as follows. In Sec.~\ref{sec:nonrel} we
derive the apparent transverse velocity of a distant radio source
in the non-relativistic approximation. In Sec.~\ref{sec:spec} we
present the corresponding derivation within the framework of
special relativity. Finally, in Sec.~\ref{sec:gen} we present a
derivation based on general relativity. In Sec.~\ref{sec:est} we
discuss how to estimate the apparent velocity of a source from its
proper motion and redshift and derive a relation for the velocity
valid for the currently favored model of the universe. We conclude
in Sec.~\ref{sec:conc}.

\section{Non-relativistic approximation}
\label{sec:nonrel}

Equation~(\ref{eq:v_a_stat}) is valid for a source that is at rest
with respect to an observer. However, extragalactic superluminal
sources participate in the global expansion of the universe.
Consequently, they have Hubble velocities of recession
proportional to their distance. For distant sources these
velocities are a considerable fraction of the speed of light. In
this section we will investigate the effect the recession velocity
of a source has on its apparent transverse velocity, in the
non-relativistic limit.

We assume that at time $t^{(1)} = 0$ a blob of matter is
ejected from the core of a radio source at an angle $\theta$ to the
line of sight of an observer located at a distance $r$ from the
source. The observer notices this event at time $t_o^{(1)} =
r/c$. At time $t^{(2)} = \Delta t$, the blob has moved to a
distance $v \Delta t$ away from the source, where $v$ is its
velocity relative to the source. The transverse displacement of the
blob from the core is then $\Delta y = v\sin \theta\Delta t$. If
the source were stationary, this motion would be noticed by the
observer at time
$t_o^{(2)} = \Delta t + r/c - (v \cos\theta \Delta t)/c$, because
at time $t^{(2)}$ the blob is closer to the observer by the distance
$v \cos\theta \Delta t$. However, due to the expansion of the
universe, the source is receding from the observer with the Hubble
velocity
$v_{\rm H}$. We first assume that both $v$ and $v_{\rm H}$
are much less than the velocity of light. Then,
\begin{equation}
t_o^{(2)} = \Delta t + \frac{r + \Delta r}{c} - \frac{v \cos\theta
\Delta t}{c},
\label{eq:t2}
\end{equation}
where $\Delta r = v_{\rm H} \Delta t$. In Eq.~(\ref{eq:t2}), we have
implicitly used the Galilean transformation of velocities, that is,
the blob velocity relative to the observer is a sum of the source
velocity relative to the observer and the blob velocity relative to
the source. Hence, the relation between the observed and
true time intervals, respectively $\Delta t_o$ and $\Delta t$, 
from the moment of emanation of the blob from the core, is
\begin{equation}
\Delta t_o \equiv t_o^{(2)} - t_o^{(1)}= (1 + \beta_{\rm H} - 
\beta \cos\theta) \Delta t,
\label{eq:Dt_0_non-rel}
\end{equation}
where $\beta_{\rm H} = v_{\rm H}/c$, and the apparent transverse
velocity of the blob measured by the observer is
\begin{equation}
v_a = \frac{\Delta y}{\Delta t_o} =
\frac{v \sin\theta}{1 + \beta_{\rm H} - \beta\cos\theta} .
\label{eq:v_app}
\end{equation}
For $\beta_{\rm H} =0$, Eq.~(\ref{eq:v_app}) reduces to
Eq.~(\ref{eq:v_a_stat}), which is valid for a blob ejected from a
{\em stationary\/} source.

{}From Eq.~(\ref{eq:v_app}) it is clear that the recession of the
source (away from the observer) opposes the effect of the blob motion
(toward the observer), in a sense that it lowers the value of the
apparent transverse velocity of the blob. For nonrelativistic
recession we have $v_{\rm H} = c z$, where $z$ is the redshift of the
source. This equation yields $\beta_{\rm H} = z$, and hence
\begin{equation}
\beta_a = \frac{\beta \sin\theta}{1 + z - \beta\cos\theta} .
\label{eq:beta_app}
\end{equation}
We reiterate that Eq.~(\ref{eq:beta_app}) is valid only when both
$z$ and $\beta$ are much less than unity. 

\section{Special-relativistic approach}
\label{sec:spec}
To appear superluminal, the velocity of a blob must be
relativistic. In this section we apply special relativity to
derive a formula for the apparent velocity of a blob, moving away from
a core located at a redshift $z$. According to special relativity,
the Cartesian components of velocities measured in the reference
frames $O$ and
$O'$, moving with constant relative velocity $V$ along the $x$-axis,
are related by
\begin{subequations}
\begin{eqnarray}
v_x &=& \frac{v_x' + V}{1 + v_x' V/c^2} 
\label{eq:v_x} \\
v_y &=& \frac{v_y'}{\gamma (1 + v_x' V/c^2)},
\label{eq:v_y}
\end{eqnarray}
\end{subequations}
where $\gamma = [1 - (V/c)^2]^{-1/2}$. Unprimed (primed)
quantities refer to the velocity components measured in the frame $O$
($O'$). We align the $x$-axis along the observer's
line of sight to the radio source. The unprimed frame is the
observer's reference frame; the primed one is the source's frame.
Then, $v_x' = - v'\cos\theta$ and $v_y' = v'\sin\theta$, where
$v'$ is the blob velocity relative to the source. The source
velocity relative to the observer is $V = v_{\rm H}$. Hence, the
Cartesian components of the blob velocity relative to the observer
are
\begin{subequations}
\label{eq:vxy2}
\begin{eqnarray}
v_x &=& \frac{v_{\rm H} - v'\cos\theta}{1 - 
\beta'\beta_{\rm H}\cos\theta}
\label{eq:v_x2} \\
v_y &=& \frac{v'\sin\theta}{\gamma_{\rm H}(1 - \beta'\beta_{\rm
H}\cos\theta)},
\label{eq:v_y2}
\end{eqnarray}
\end{subequations}
where $\gamma_{\rm H} = (1 - \beta_{\rm H}^2)^{-1/2}$. 

In Eqs.~(\ref{eq:vxy2}), the time dilation effect is naturally
accounted for. However, in the observer's inertial frame we have to
take into account the extra time dilation factor that occurs because
the distance to the emitting blob (and thus the distance light has to
propagate to reach the observer) is changing. In the time $\Delta t$
the emitter moves a distance $v_x \Delta t$ away from the observer
($v_x$ and the distance may be negative). The total observed time,
$\Delta t_o$, is $\Delta t$ plus the extra factor describing how long
it takes light to traverse this extra distance ($v_x \Delta t/c$),
\begin{equation}
\Delta t_o = \Delta t + \frac{v_x \Delta t}{c} = (1 + \beta_x)
\Delta t .
\label{eq:Deltat}
\end{equation}
Note that Eq.~(\ref{eq:Deltat}) is similar to
Eq.~(\ref{eq:Dt_0_non-rel}), but instead of the non-relativistic
approximation for $v_x$, $v_x = v_{\rm H} - v'\cos\theta$, we
have used the special relativity formula~(\ref{eq:v_x2}). The
transverse distance covered by the blob in the time $\Delta t$ is
$\Delta y = v_y
\Delta t$. Hence, the apparent transverse velocity is $v_a = \Delta
y/\Delta t_o = v_y/(1 +
\beta_x)$. From Eq.~(\ref{eq:vxy2}) we have
\begin{equation}
v_a = \frac{v'\sin\theta}{\gamma_{\rm H}(1 - \beta'\beta_{\rm
H}\cos\theta)} \Big(1 + \frac{\beta_{\rm H} -
\beta'\cos\theta}{1 - \beta'\beta_{\rm H}\cos\theta}\Big)^{-1}, 
\label{eq:v_abs}
\end{equation}
or, 
\begin{equation}
\beta_a = \frac{\gamma_{\rm H}^{-1} \beta'
\sin\theta}{(1+\beta_{\rm H}) (1 - \beta'\cos\theta)} .
\label{eq:beta_obs}
\end{equation}
Next, we have
\begin{equation}
\frac{\gamma_{\rm H}^{-1}}{1+\beta_{\rm H}} = \frac{(1 - \beta_{\rm
H}^2)^{1/2}}{1+\beta_{\rm H}} = \left(\frac{1 - \beta_{\rm H}}{1 + 
\beta_{\rm H}} \right)^{1/2} = (1 + z)^{-1} .
\label{eq:gamma} 
\end{equation}
In the last step we have used the expression 
for the special relativitistic Doppler effect. The result is
\begin{equation}
\beta_a = \frac{\beta \sin\theta}{(1 + z) (1 - \beta\cos\theta)} .
\label{eq:beta_fin}
\end{equation}
For consistency of notation, we have omitted the
primes on $\beta$ in Eq.~(\ref{eq:beta_fin}). However, it should
be clear that in Eq.~\eqref{eq:beta_fin}, as in
Eq.~(\ref{eq:beta_app}), $v = \beta c$ is the blob velocity {\em
relative to the source}.

Equation~(\ref{eq:beta_fin}) shows that the recession (Hubble)
velocity of a radio source reduces the amplitude of the apparent
transverse motion of its blob by the factor of $1 + z$. When both
$z$ and $\beta$ are much less than unity, the term $z \beta
\cos\theta$ can be neglected and Eq.~(\ref{eq:beta_fin}) reduces to
its non-relativistic limit, Eq.~(\ref{eq:beta_app}). What is the range
of applicability of Eq.~(\ref{eq:beta_fin})? In its derivation we have
not made any explicit assumptions about the values of $z$ and $\beta$,
except that $z \ge 0$ and $\beta\in[0,1]$. However, when applied to
the real world, special relativity is guaranteed to work only locally,
because, in general, it is not possible to eliminate the gravitational
field globally by a suitable choice of a reference frame. (There are
no global inertial frames in the universe.) Thus, special relativity
applies only to a limited region around a radio source. Consequently,
though it has not been assumed explicitly, Eq.~(\ref{eq:beta_fin}) is
valid for arbitrary $\beta\in[0,1]$, but only for $z \ll 1$. To find
its generalization for arbitrary $z$, we need to apply general
relativity.

\section{General-relativistic approach}
\label{sec:gen}
The metric of a homogeneous and isotropic universe is given by the
Robertson--Walker line element:
\begin{equation}
c^2 ds^2 = c^2 dt^2 - a^2(t)[d\chi^2 + R_0^2 S^2(\chi/R_0) 
(d\theta^2 + \sin^2\theta d\phi^2)] .
\label{eq:RW}
\end{equation}
Here, $R_0^{-2}$ is the curvature of the universe and the function
$S(x)$ equals $\sin(x)$, $x$, and $\sinh(x)$ for a
closed, flat, and open universe, respectively. The function $a(t)$
is called a scale factor and relates the physical, or proper,
coordinates of a galaxy,
$\bfr$, to its fixed or comoving coordinates, $\mathbf{\chi}$:
$\bfr = a \mathbf{\chi}$. This function accounts for the expansion
of the universe; its detailed time dependence is determined by
the Friedman equations.\cite{friedman} We normalize $a$
so that at the present time, $a(t_0) = 1$.

Photons propagate along null geodesics, $ds = 0$. If we place an
observer at the origin of the coordinate system, the geodesic of
the photons emitted by the source toward the observer is radial.
The source's comoving radial coordinate is $\chi$. From the
metric~(\ref{eq:RW}), we have for the photons emitted from the core
\begin{equation}
\int_{t_e^{(1)}}^{t_o^{(1)}} \frac{c dt'}{a(t')} = 
\!\int_0^\chi d\chi' = \chi .
\label{eq:source}
\end{equation}
For the photons emitted later from the blob,
\begin{equation}
\int_{t_e^{(2)}}^{t_o^{(2)}} \frac{c dt'}{a(t')} = \chi - \Delta
\chi,
\label{eq:blob}
\end{equation}
where $\Delta \chi$ is the comoving distance the blob has covered,
projected on the line of sight. Its relation to the proper
distance, $\Delta l$, is $\Delta l = a(t_e) \Delta \chi = (1 +
z)^{-1}
\Delta \chi$, where $z$ is the source's redshift. The proper
distance is $v \cos\theta \Delta t_e$. Hence
\begin{equation}
\Delta \chi = (1 + z) v \cos\theta \Delta t_e .
\label{eq:Dchi}
\end{equation}
The subtraction of Eq.~(\ref{eq:blob}) from Eq.~(\ref{eq:source})
yields
\begin{equation}
\Delta \chi = \! \left[\int_{t_e^{(1)}}^{t_o^{(1)}} -
\int_{t_e^{(2)}}^{t_o^{(2)}} \right] \frac{c dt'}{a(t')} = 
\!\! \int_{t_e^{(1)}}^{t_e^{(2)}} \frac{c dt'}{a(t')} - 
\int_{t_o^{(1)}}^{t_o^{(2)}} \frac{c dt'}{a(t')} .
\label{eq:Dchi_int} 
\end{equation}
Because both $\Delta t_e$ and $\Delta t_o$ are very small compared
to the Hubble time, Eq.~(\ref{eq:Dchi_int}) simplifies to
\begin{equation}
\Delta \chi + \frac{c \Delta t_o}{a(t_o)} = 
\frac{c \Delta t_e}{a(t_e)},
\label{eq:Dchi_diff} 
\end{equation}
or
\begin{equation}
\Delta \chi + c \Delta t_o = (1 + z) c \Delta t_e .
\label{eq:Dchi_diff2} 
\end{equation}
By using Eq.~(\ref{eq:Dchi}), we obtain
\begin{equation}
\Delta t_o = (1 + z) \Delta t_e (1 - \beta \cos\theta) .
\label{eq:Dt_o} 
\end{equation}
For vanishing blob velocity, Eq.~(\ref{eq:Dt_o}) describes the
well-known phenomenon of cosmological time dilation. The transverse
component of the distance covered by the blob is $\Delta y = v
\sin\theta \Delta t_e$. The apparent transverse velocity is $\Delta
y/\Delta t_o$, and hence
\begin{equation}
\beta_a = \frac{\beta \sin\theta}{(1+z) (1 - \beta\cos\theta)},
\label{eq:beta_final} 
\end{equation}
in agreement with Eq.~(\ref{eq:beta_fin}). Thus, in this 
case, the prediction of special relativity turns out to be valid
globally, that is, for arbitrary $z$. Why?

The amplitude of the apparent transverse motion of a distant radio
source is reduced by a factor that depends only on its redshift; it
does not depend on the background cosmological model. In particular,
it does not depend on the mean densities of the different energy
components in the universe, $\Omega_\Lambda$ and $\Omega_m$. Here,
$\Omega_\Lambda$ denotes the cosmological constant $\Lambda$ expressed
in units of the critical density, and $\Omega_m$ is the mean density
of non-relativistic matter in the universe, also in units of the
critical density. Physically, these components cause the universe's
acceleration and deceleration, respectively. Therefore, the lack of
sensitivity of the reduction factor (of the apparent transverse
velocity) to their densities implies that the factor is insensitive to
the acceleration or deceleration of the cosmological expansion, so its
origin is kinematic. Mathematically, the reduction is the same for any
$\Omega_\Lambda$ and $\Omega_m$, so in particular for $\Omega_\Lambda
= \Omega_m = 0$, that is, for an empty universe. This particular
cosmological model is called the Milne model, or {\em kinematic
cosmology}. The dynamics of an empty universe can be completely
described by special relativity,\cite{longair} which is why its
prediction turns out to be valid globally here. However, this
conclusion holds only a posteriori, that is after applying general
relativity and finding out that the reduction factor depends only on
the redshift of the source.

Quantitatively, the amplitude of the apparent transverse motion of a
radio source located at a redshift $z$ is reduced by a factor $1 +
z$. The general relativistic explanation of this fact is the
cosmological time dilation between the observer's frame and the
frame of the object.\cite{pz87} The special relativistic
explanation is the recession velocity of the object, resulting in
the same amount of time dilatation. These explanations are mutually
consistent, because the origin of cosmological time dilation is the
expansion of the universe, which causes distant galaxies to recede
from the Milky Way. We have argued that the reduction of the
apparent velocity is in essence a kinematic effect and can be
qualitatively explained within a non-relativistic framework, as
demonstrated in Sec.~\ref{sec:nonrel}. Relativistic corrections are
necessary to describe the effect quantitatively.

\section{Estimating the apparent velocity}
\label{sec:est}

The physical size of an object at the redshift $z$ that subtends the
angle $\Delta \phi$ on the sky, $\Delta y$, can be readily derived
from the metric in Eq.~(\ref{eq:RW}). The result is
\begin{equation}
\Delta y = D_A \Delta \phi.
\label{eq:Dy} 
\end{equation}
Hence, the
apparent transverse velocity of a radio source is
\begin{equation}
v_a = \mu D_A,
\label{eq:v_a} 
\end{equation}
where $\mu = \Delta \phi/ \Delta t_o$ is its observed proper
motion. By measuring the redshift and the proper motion of a radio
source and knowing the cosmological parameters, we can estimate its
apparent velocity. This estimate can be subsequently used to constrain
the combination of the parameters characterizing the source, given by
the right-hand side of Eq.~(\ref{eq:beta_final}).  To constrain only
the internal parameters of the source that describe its kinematics and
geometry (that is, $v$ and $\theta$), it is necessary to eliminate the
dependence of $v_a$ on the redshift. Therefore, instead of $v_a$
itself, it is common to estimate the quantity
\begin{equation}
v_m = (1 + z) v_a .
\label{eq:v_m} 
\end{equation} 

Unfortunately, $v_m$ is widely called the ``apparent velocity'' (see,
for example, Refs.~\onlinecite{pz87,porc,zp88,ver94,jorst}).  This
term is misleading, because it disguises the fact that the true
apparent velocity, $v_a$, is affected by the recession velocity of the
source. We will therefore distinguish between $v_a$ and $v_m$, and
will call the latter the ``velocity measure.'' The latter is the
velocity that an observer would measure if he/she were located at the
redshift $z$ in the vicinity of the source. This velocity is not the
velocity we measure on Earth. We have
\begin{equation}
v_m = (1 + z) \mu D_A = \mu D,
\label{eq:v_m2} 
\end{equation}
where $D = (1+z) D_A$ is the distance measure. We recall that an
object of luminosity $L$ has flux $f = L/(4\pi D_L^2)$, where
\begin{equation}
D_L = (1+z) D = (1+z)^2 D_A
\label{eq:D_L} 
\end{equation} 
is the luminosity distance.

To date, all analyses of superluminal motion in extragalactic radio
sources have assumed that the universe has a vanishing cosmological
constant at low redshifts dominated by non-relativistic
matter.\cite{except} For such a universe, Mattig\cite{mattig} derived
an analytical expression for the distance measure $D$ for arbitrary
$\Omega_m$. Consequently, the classical Mattig formula for $D$ has
been used in all analyses of superluminal radio sources (see, for
example, Refs.~\onlinecite{pz87,porc,cohen,ver94,jorst}). However,
observations in cosmology consistently imply that $\Lambda$ is about
0.7 in units of the critical density (that is, $\Omega_\Lambda \simeq
0.7$). The distance measure $D$ (or the luminosity distance $D_L$) is
sensitive to the presence of $\Lambda$. Thanks to this sensitivity,
the Hubble diagram for supernovae Ia has been successfully used to
show that the universe has a non-zero $\Lambda$. Therefore, we need a
relation for the distance that accounts for the presence of $\Lambda$.

Observations also strongly suggest that the universe must be very
close to spatially flat (see, for example, Ref.~\onlinecite{wmap}).
For a flat universe with a cosmological constant, the luminosity
distance is (see, for example, Ref.~\onlinecite{peacock})
\begin{equation}
D_L(z) = c H_0^{-1} (1 + z) \! \int_0^z \! \left[1 - \Omega_m +
\Omega_m (1 + x)^3 \right]^{-1/2} {\rm d}x .
\label{eq:DL_int}
\end{equation} 
Here, $H_0$ denotes the Hubble constant, and spatial flatness requires
$\Omega_\Lambda = 1 - \Omega_m$. For small redshifts, $z \ll 1$,
Eq.~(\ref{eq:DL_int}) reduces to $D_L = c H_0^{-1} z$. The integral in
Eq.~(\ref{eq:DL_int}) cannot be performed analytically; it is a
Weierstrass elliptic function.\cite{ds87}

Pen\cite{pen} derived an approximate analytic expression for
$D_L(z)$, accurate to better than 0.4\% for $0.2 \le
\Omega_m \le 1$, for any redshift. The value of $\Omega_m$ is
currently known to be around 0.3. Pen's
approximation for the luminosity distance is
\begin{equation}
D_L = c H_0^{-1} (1 + z) \big[\eta(0,\Omega_m) -
\eta(z,\Omega_m) \big],
\label{eq:DL}
\end{equation}
where 
\begin{eqnarray}
\eta(z,\Omega_m) &=& 2 \Omega_m^{-1/2} 
\big[ (1+z)^4 - 0.1540 (1+z)^3 s 
\nonumber \\
&&{}+ 0.4304 (1+z)^2 s^2 + 
0.19097\, (1+z) s^3 
+ 0.066941 s^4 \big]^{-1/8} ,
\label{eq:eta}
\end{eqnarray}
and
\begin{equation}
s = \Big(\frac{1 - \Omega_m}{\Omega_m} \Big)^{1/3} .
\label{eq:s}
\end{equation} 
(The original notation has been slightly modified.) It is a
matter of choice whether to use a simple numerical integral or
its fairly complex analytic approximation. If one chooses the latter
approach, then from Eqs.~(\ref{eq:v_m2}) and (\ref{eq:DL}), the
velocity measure $v_m$, or $\beta_m = v_m/c$, is
\begin{equation}
\beta_m = \mu H_0^{-1} \left[\eta(0,\Omega_m) -
\eta(z,\Omega_m) \right],
\label{eq:beta_m}
\end{equation}
with $\eta$ given by Eq.~(\ref{eq:eta}). Note that $\beta_m$ is
dimensionless, as it should be. Equation~(\ref{eq:v_m}) implies that
\begin{equation}
\beta_a = \mu H_0^{-1} (1+z)^{-1} \left[\eta(0,\Omega_m) -
\eta(z,\Omega_m) \right] .
\label{eq:beta_o}
\end{equation}

For completeness, we provide current estimates of the Hubble
constant and $\Omega_m$. From a joint analysis of the SDSS and the
WMAP data, Tegmark et al.\cite{teg04} deduced that $H_0 =
70^{+4}_{-3}\, \kms
\cdot \hbox{Mpc}^{-1}$, and $\Omega_m = 0.30 \pm 0.04$ ($68$\%
confidence intervals). 

\section{Conclusions}
\label{sec:conc}
The apparent transverse velocity of a radio source located at a
redshift $z$ is suppressed by a factor $1 + z$. We have derived
this result within the framework of both
special and general relativity. The underlying cause of this
suppression is the recession of the source due to the expansion of
the universe.

Given the values of the Hubble constant and the mean densities of
different energy components in the universe, the apparent velocity
of a source can be estimated from its redshift and proper motion.
We have derived a relation for the velocity valid for the currently
favored cosmological model, that is, a flat universe with a
non-zero cosmological constant.

\begin{acknowledgments}

I am thankful to Marek Sikora for stimulating discussions during the
preparation of the manuscript. Two anonymous referees are warmly
acknowledged for their apt comments and helpful suggestions concerning
an earlier version of this paper. The editor of AJP has considerably
amended the text. This research has been supported in part by the
Polish State Committee for Scientific Research grant No.~1 P03D 012
26, allocated for the period 2004--2007.
\end{acknowledgments}


\begin{thebibliography}{99}
\bibitem{redshift} In other words, redshift is a shift in the 
frequency of photons toward lower energy, or longer wavelength (i.e.,
the red part of the electromagnetic spectrum). The more distant is the
source, the higher is its redshift.
\bibitem{cite} ``To derive the apparent transverse velocity
$\beta_{\rm app}$ from an observed proper motion $\mu$ it is
necessary to take account of the time-dilation between the
observer's frame and the frame of the object: $\beta_{\rm app} =
\mu D_A (1 + z)/c$.''\cite{pz87} $D_A$ is the angular diameter 
distance and the extra factor of $1 + z$ is due to time dilation. If
we combine this relation with the relation between the angular
diameter distance and the luminosity distance, Eq.~(\ref{eq:D_L}), and
a non-standard form of the Mattig formula for the
latter,\cite{non-st} we obtain an expression for the apparent
transverse velocity (Eq.~(4) of Ref.~\onlinecite{pz87}), repeatedly
used in the literature.
\bibitem{lambda} The cosmological constant, $\Lambda$, is a quantity
occurring in Einstein's theory of general relativity. It is
proportional to the energy density of the vacuum. The constant can be
thought of as the amount of energy that is embedded in empty space. If
it is positive then the expansion of space would release more energy
that in turn would accelerate the expansion, whereas if it is negative
the expansion of space would consume energy. Recent observations
indicate that the expansion of the universe is accelerating and if
this accleration is due entirely to the cosmological constant, then
its value is on the order of $10^{-35}{\rm sec}^{-2}$, or about $0.7$
in units of the critical density.\cite{value}
\bibitem{except} An exception is Ref.~\onlinecite{keller}, where
it was assumed that $\Omega_\Lambda = 0.7$. However, no details are
given of the calculation of the apparent velocity from the proper
motion.
\end{thebibliography}
\end{document}